\documentclass[AMA,STIX2COL]{WileyNJD-v2}
\usepackage{moreverb}
\usepackage{csquotes}
\usepackage{graphicx,color}
\usepackage{float}
\usepackage[caption=false]{subfig}
\usepackage{amsmath}
\usepackage{amsmath}
\usepackage{multirow}
\usepackage{dsfont}
\usepackage[shortlabels]{enumitem}
\usepackage{epstopdf}
\newcommand\BibTeX{{\rmfamily B\kern-.05em \textsc{i\kern-.025em b}\kern-.08em
T\kern-.1667em\lower.7ex\hbox{E}\kern-.125emX}}
\articletype{ORIGINAL ARTICLE}%
\received{<day> <Month>, <year>}
\revised{<day> <Month>, <year>}
\accepted{<day> <Month>, <year>}
\begin{document}
\title{Kinetic contribution to the arbitrary order odd frequency moments of the dynamic structure factor}
\author{Panagiotis Tolias$^1$}
\author{Tobias Dornheim$^{2,3}$}
\author{Jan Vorberger$^3$}
\address{$^1$\orgdiv{Space and Plasma Physics}, \orgname{Royal Institute of Technology (KTH)}, \orgaddress{\state{Stockholm}, \country{Sweden}}}
\address{$^2$\orgname{Center for Advanced Systems Understanding (CASUS)}, \orgaddress{\state{G\"orlitz}, \country{Germany}}}
\address{$^3$\orgname{Helmholtz-Zentrum Dresden-Rossendorf (HZDR)}, \orgaddress{\state{Dresden}, \country{Germany}}}
\authormark{P. Tolias}
\corres{Panagiotis Tolias, Space and Plasma Physics, Royal Institute of Technology, Stockholm SE-100 44, Sweden. \email{tolias@kth.se}}
\abstract[Abstract]{An exact expression is derived for the kinetic contribution to the odd (arbitrary order) frequency moments of the dynamic structure factor via a finite summation that features averages of even (all lower orders) powers of the momentum over the exact momentum distribution. The derivation is carried out for the non-interacting Fermi gas and generalized to the interacting case based on the conjecture that averages over the Fermi distribution can be substituted with averages over the exact distribution. The expression is validated against known results (first, third frequency moments) and new explicit calculations (fifth, seventh frequency moments).}
\keywords{frequency moments; linear response theory; non-interacting electron gas; correlated quantum systems}
\maketitle

\section{Introduction}\label{sec:intro}

The concrete understanding of interacting quantum many-body systems at finite temperatures remains a formidable challenge in condensed matter physics, quantum chemistry and plasma physics\,\cite{stefanucci2013nonequilibrium,bonitz_book,abrikosov_book}. This holds particularly true for warm dense matter (WDM), an extreme state of matter that is characterized by the complex interplay between electronic quantum effects (exchange and diffraction), partial ionization, Coulomb interactions and thermal excitations\,\cite{Remington_2005,GrazianiBook,RileyBook}, see also the recent dedicated roadmap by Vorberger \emph{et al.}\,\cite{vorberger2025roadmapwarmdensematter}. Despite its exotic nature, WDM naturally emerges in dense astrophysical objects (giant gas planet interiors\,\cite{Howard2023,Hernandez2023,Breuer2023}, white dwarf envelopes\,\cite{Kritcher2020,Saumon2022}, neutron star crusts\,\cite{chamel_lrr_2008,neutronstar_2021}) and is also highly relevant in the early compression stage of the pellet during inertial confinement fusion\,\cite{Hurricane_RevModPhys_2023,hu_ICF}. A pre-requisite for the understanding of WDM is the quantitative description of the physics of the finite temperature uniform electron gas (UEG), the archetypal model system of interacting electrons where the charge neutralizing ions are assumed to act as a rigid uniform background\,\cite{Baus_Hansen_OCP,plasma2,IIT,DornheimPhysRep2018}. In the thermodynamically stable case of equal spin-up and spin-down electrons, the UEG states are uniquely specified by the quantum coupling parameter $r_{\mathrm{s}}=d/a_{\mathrm{B}}$ (ratio of the Wigner-Seitz radius over the Bohr radius) and the degeneracy parameter $\Theta=T/E_{\mathrm{f}}$ (ratio of the thermal energy over the Fermi energy). In WDM conditions, there is an absence of small parameters ($r_{\mathrm{s}}\sim\Theta\sim1$), which clearly indicates that a holistic treatment of the full interplay between quantum effects, Coulomb correlations and thermal excitations is required\,\cite{DornheimPhysRep2018,Ott2018,DornheimPOP2023}.

Recent breakthroughs in quantum Monte Carlo (QMC) simulations\,\cite{DornheimPhysRep2018,DornheimPOP2023,DornheimPOP2017,BonitzPOP2020,XiongPRE2023} have provided a wealth of quasi-exact data for the thermodynamic\,\cite{KarasievPRL2014,DornheimPRL2016,GrothPRL2017,KarasievPRL2018,DornheimPRBL2025,DornheimPRR2025,DornheimPRB2025}, structural\,\cite{DornheimJCP2019ESA,DornheimPRL2020,DornheimPRB2021ESA,DornheimPRR2022spin,DornheimJPCL2024}, dynamic\,\cite{DornheimPRL2018,GrothPRB2019,HamannPRB2020,DornheimJCP2023xi,ChunaJCP2025} and non-linear\,\cite{DornheimPRL2020nonlin,DornheimPRR2021nonlin,DornheimJCP2021nonlin,ToliasEPL2023} properties of the finite temperature uniform electron gas. Even outside the realm of QMC simulations, there remains a wealth of exact UEG results which mainly concern the long-wavelength plasmon-dominated behavior\,\cite{ChunaJCP2025,PinesIBook,KuglerPRA1970}, the short-wavelength quantum-dominated behavior\,\cite{NiklassonPRB1974,HolasCollection1987,VignalePRB1988,KimballJPA1975,YasuharaPA1976,HofmannPRB2013,RajagopalPRB1978} and the high-frequency behavior\,\cite{PinesIBook,HolasCollection1987,GiulianiVignale,KuglerJSP1975}. In particular, exact high-frequency limits are tightly connected with the frequency moments of spectral functions. Within the framework of linear response theory, the latter are expressed as equilibrium expectation values of equal-time operators that are known as sum rules\,\cite{GiulianiVignale}. The frequency moments of a spectral function $A(\boldsymbol{q},\omega)$ are defined by\,\cite{GiulianiVignale,TkachenkoBook}
\begin{equation}
M^{(\alpha)}_{\mathrm{A}}(\boldsymbol{q})=\langle\omega^{\alpha}\rangle_{\mathrm{A}}=\int_{-\infty}^{+\infty}d\omega\;\omega^{\alpha}A(\boldsymbol{q},\omega)\,,\label{momentsgen}
\end{equation}
with $\alpha$ an arbitrary positive or negative integer. Spectral functions of particular interest are the imaginary part of the linear density response $\Im\{\chi(\boldsymbol{q},\omega)\}$, the dynamic structure factor $S(\boldsymbol{q},\omega)$ and the dielectric loss function $-\Im\{\epsilon^{-1}(\boldsymbol{q},\omega)\}/\omega$. The most fundamental is $\Im\{\chi(\boldsymbol{q},\omega)\}$ that serves as the linear response theory generator of odd frequency moments\,\cite{GiulianiVignale}.

Modern theoretical research centered around the frequency moments remains particularly active. Very recently, Tolias \emph{et al.} derived an exact infinite series representation for the otherwise unknown even frequency moments of the dynamic structure factor that expresses them through the known odd frequency moments of the dynamic structure factor as well as proposed and validated truncated series versions that allow for the accurate evaluations of the explicitly unknown second, fourth, fifth frequency moments for the weakly degenerate finite temperature UEG\,\cite{ToliasPOP2025}. Moreover, Dornheim \emph{et al.} introduced an exact framework that allows the direct extraction of the even and odd frequency moments of the dynamic structure factor from imaginary-time (density-density) correlation functions evaluated in QMC simulations without the need for analytic continuation and provided quasi-exact results for the first five frequency moments of the dynamic structure factor of the finite temperature UEG\,\cite{DornheimPRBmom}. Finally, Filinov \emph{et al.} proposed an advanced nine moment variant of the non-perturbative sum rule approach (which reconstructs the loss function from a finite set of its frequency moments on the basis of the Nevanlinna theorem) that determines the elusive sixth and eighth frequency moments from a two-parameter Shannon information entropy maximization procedure combined with a static approximation for a dynamic parametric function and demonstrated the accuracy of the predicted dynamic UEG properties against QMC simulations utilizing input for the static inverse dielectric function and the static structure factor\,\cite{FilinovPRB2023}.

In this article, we derive an exact expression for the kinetic contribution to the arbitrary order odd ($2k+1$) frequency moments of the dynamic structure factor of the finite temperature paramagnetic UEG through a finite summation that features the thermodynamic averages of the even powers (from $0$ up to $2k$) of the momentum magnitude over the exact momentum distribution. The arbitrary order derivation begins with the non-interacting Fermi gas, whose known odd frequency moments are manipulated so that the Fermi distribution explicitly emerges and so that Fermi distribution averages of even powers of the momentum arise. Then, it is conjectured that the non-interacting result remains the same in the general interacting case, provided that the thermodynamic average is over the exact momentum distribution instead of the Fermi distribution. The ansatz-based calculations are validated against known results (first, third frequency moments) and new explicit calculations (fifth, seventh frequency moments). The generality of the expression and its significance are also discussed.

\section{Theoretical background}\label{sec:backgroumd}

\subsection{Odd frequency moments for the dynamic structure factor}\label{sec:backgroumdSumRules}

The general expression for the odd frequency moments of the imaginary part of the linear response function constitutes one of the most profound results of linear response theory\,\cite{GiulianiVignale,KuboBook}. The expression stems from the high frequency expansion of the Kramers-Kronig expression combined with the short time expansion of the Kubo formula. In its density--density version, the sum rule expression connects the odd frequency moments of the imaginary part of the linear density response function to the equilibrium expectation value of an equal-time commutator\,\cite{GiulianiVignale,IchimaruBook,SingwiSSP1981}. The fundamental expression can be written in a differential form that only involves the microscopic density operator $\hat{\rho}$\,\cite{IchimaruBook}
\begin{equation*}
{M}_{\mathrm{Im}\chi}^{(2k+1)}(\boldsymbol{q})=(-1)^{k+1}\frac{\pi}{V}\frac{\imath}{\hbar}\left\langle\left[\left.\frac{\partial^{2k+1}\hat{\rho}(\boldsymbol{q},t)}{\partial{t}^{2k+1}}\right|_{t=0},\hat{\rho}^{\dagger}(\boldsymbol{q})\right]\right\rangle_0\,,
\end{equation*}
with $V$ the volume and the subscript "$0$" referring to the unperturbed system. The fundamental expression can be rewritten, after repeated differentiations and applications of the Heisenberg equation of motion, in a nested commutator form that only involves the microscopic density operator $\hat{\rho}$ and the Hamiltonian operator $\hat{H}$\,\cite{GiulianiVignale}
\begin{align*}
{M}_{\mathrm{Im}\chi}^{(2k+1)}(\boldsymbol{q})=(-1)^{k+1}\frac{\pi}{V}\left(\frac{\imath}{\hbar}\right)^{2k+2}&\left\langle\left[\left[\hat{H},\left[\hat{H}\cdots[\hat{H},\hat{\rho}(\boldsymbol{q})]\cdots\right]\right],\right.\right.\nonumber\\\quad&\left.\left.\hat{\rho}^{\dagger}(\boldsymbol{q})\right]\right\rangle_0\,,
\end{align*}
where the number of $\hat{\rho}(\boldsymbol{q})-$bound Hamiltonian operators is $2k+1$. The fundamental expression can also be rewritten, after exploiting the fact that the thermodynamic equilibrium average of equal-time commutators cannot depend on the time instant given the invariance under time translation, in a distributed nested commutator form that is the most convenient for explicit evaluations of the frequency moment sum rules\,\cite{KuglerJSP1975}
\begin{align}
{M}_{\mathrm{Im}\chi}^{(2k+1)}(q)=(-1)^{k+\ell}\frac{\pi}{V}&\left(\frac{\imath}{\hbar}\right)^{2k+2}\left\langle\left[\left[\cdots\left[\left[\hat{\rho}(\boldsymbol{q}),\hat{H}\right],\hat{H}\right]\cdots\right],\right.\right.\nonumber\\\quad&\left.\left.\left[\cdots\left[\left[\hat{\rho}^{\dagger}(\boldsymbol{q}),\hat{H}\right],\hat{H}\right]\cdots\right]\right]\right\rangle_0\,.\label{eq:sumruleImchi}
\end{align}
where $\ell\leq2k+1$ is a non-negative integer that is at our disposal, with $2k+1-\ell$ the number of $\hat{\rho}(\boldsymbol{q})-$bound Hamiltonian operators and $\ell$ the number of $\hat{\rho}^{\dagger}(\boldsymbol{q})-$bound Hamiltonian operators. For the sake of completeness, we mention that the even frequency moments of the imaginary part of the linear density response are all zero from the frequency moment definition and the $\Im\{\chi(q,-\omega)\}=-\Im\{\chi(q,\omega)\}$ frequency parity\,\cite{GiulianiVignale}.

The definition of the frequency moments, the fluctuation--dissipation theorem and the $\Im\{\chi\}$ frequency parity also yield an algebraic correspondence rule between the odd frequency moments of the dynamic structure factor and the odd frequency moments of the imaginary part of the linear density response function\,\cite{KuglerJSP1975}. The correspondence rule reads as\,\cite{ToliasPOP2025,DornheimPRBmom}
\begin{equation}
{M}_{\mathrm{S}}^{(2k+1)}(\boldsymbol{q})=-\frac{\hbar}{2\pi{n}}{M}_{\mathrm{Im}\chi}^{(2k+1)}(q)\,,\label{eq:correspondencerule}
\end{equation}
with $n$ the density. Thus, the odd frequency moment sum rules for the dynamic structure factor can now be obtained simply by combining Eqs.(\ref{eq:sumruleImchi},\ref{eq:correspondencerule}). This leads to the fundamental expression
\begin{align}
{M}_{\mathrm{S}}^{(2k+1)}(\boldsymbol{q})=(-1)^{k+\ell+1}&\frac{\hbar}{2{N}}\left(\frac{\imath}{\hbar}\right)^{2k+2}\left\langle\left[\left[\cdots\left[\left[\hat{\rho}(\boldsymbol{q}),\hat{H}\right],\hat{H}\right]\cdots\right],\right.\right.\nonumber\\\quad&\left.\left.\left[\cdots\left[\left[\hat{\rho}^{\dagger}(\boldsymbol{q}),\hat{H}\right],\hat{H}\right]\cdots\right]\right]\right\rangle_0\,,\label{eq:sumruleDSF}
\end{align}
with $N$ the particle number. In the iterated commutator structure of the odd frequency moment sum rules, the total number of Hamiltonian nests is equal to the frequency moment order. This implies that explicit calculations are always cumbersome and promptly become prohibitive, even for the simplest interacting system such as the UEG. As a consequence, in the finite temperature quantum case, only the first frequency moment (f-sum rule)\,\cite{PlaczekPR1952} and the third frequency moment (cubic sum rule)\,\cite{PuffPRA1965} have been explicitly computed. In the thermodynamic limit, they respectively read as\,\cite{GiulianiVignale,KuglerJSP1975,IchimaruBook,SingwiSSP1981}
\begin{align}
M&^{(1)}_{\mathrm{S}}(\boldsymbol{q})=\frac{\hbar{q}^2}{2m}\,,\label{eq:DSFMOM1}\\
M&^{(3)}_{\mathrm{S}}(\boldsymbol{q})=\frac{\hbar{q}^2}{2m}\left\{\left(\frac{\hbar{q}^2}{2m}\right)^2+\frac{2q^2}{m}\langle\hat{K}\rangle_0+\frac{{n}q^2}{m}U(\boldsymbol{q})+\right.\nonumber\\&\left.\frac{1}{m}\int\frac{d^3k}{(2\pi)^3}U(\boldsymbol{k})\left(\frac{\boldsymbol{k}\cdot\boldsymbol{q}}{q}\right)^2\left[S(\boldsymbol{q}-\boldsymbol{k})-S(\boldsymbol{k})\right]\right\}\,,\label{eq:DSFMOM3}
\end{align}
where $S(\boldsymbol{q})$ is the static structure factor, $U(\boldsymbol{q})=4\pi{e}^2/q^2$ is the Fourier transformed Coulomb pair interaction and $\langle\hat{K}\rangle_0$ is the kinetic energy per particle. For completeness, we point out that there are two more frequency moment sum rules that are known for the dynamic structure factor; the inverse moment that is a nontrivial consequence of the Kramers-Kronig relations and the fluctuation--dissipation theorem as well as the zeroth moment that is a trivial consequence of the Fourier transform. These respectively read as\,\cite{KuglerPRA1970,MahanBook,DornheimMRE2023}
\begin{align}
M&^{(-1)}_{\mathrm{S}}(\boldsymbol{q})=-\frac{\hbar}{2n}\chi(\boldsymbol{q})\,,\label{eq:DSFMOMinv}\\
M&^{(0)}_{\mathrm{S}}(\boldsymbol{q})=S(\boldsymbol{q})\,,\label{eq:DSFMOM0}
\end{align}
with $\chi(\boldsymbol{q})=\chi(\boldsymbol{q},\omega=0)$ the static density response function.

The pure kinetic contribution to the odd dynamic structure factor frequency moments of arbitrary order is obtained from Eq.(\ref{eq:sumruleDSF}) by considering only the kinetic part of the Hamiltonian, \emph{i.e.} setting $\hat{H}\equiv\hat{K}$. Courtesy of $[\hat{\rho}(\boldsymbol{q}),\hat{H}]=[\hat{\rho}(\boldsymbol{q}),\hat{K}]$ and $[\hat{\rho}^{\dagger}(\boldsymbol{q}),\hat{H}]=[\hat{\rho}^{\dagger}(\boldsymbol{q}),\hat{K}]$ in the non-relativistic limit, the pure kinetic contribution to the odd dynamic structure factor frequency moments of arbitrary order can be expressed as
\begin{align}
{M}_{\mathrm{S,KIN}}^{(2k+1)}(\boldsymbol{q})=(-1)^{k+\ell+1}&\frac{\hbar}{2{N}}\left(\frac{\imath}{\hbar}\right)^{2k+2}\left\langle\left[\left[\cdots\left[\left[\hat{\rho}(\boldsymbol{q}),\hat{H}\right],\hat{K}\right]\cdots\right],\right.\right.\nonumber\\\quad&\left.\left.\left[\cdots\left[\left[\hat{\rho}^{\dagger}(\boldsymbol{q}),\hat{H}\right],\hat{K}\right]\cdots\right]\right]\right\rangle_0\,,\label{eq:sumruleDSFkin}
\end{align}
where the implied operators are all kinetic energy operators.

\subsection{Frequency moments for the dynamic structure factor in the non-interacting limit}\label{sec:backgroumdFermi}

For the non-interacting Fermi gas, the linear density response function (often referred to as Lindhard response), the dynamic structure factor and the imaginary time (density--density) correlation function (ITCF) are all explicitly known\,\cite{GiulianiVignale,MahanBook,ToliasCTPP2025b}. Thus, the arbitrary order (odd and even) frequency moments of the non-interacting dynamic structure factor can be calculated in a straightforward manner. The most convenient calculation route is based on the recently-obtained connection between the arbitrary order frequency moment of the dynamic structure factor and the same order derivative of the ITCF at the imaginary time origin $\tau=0$\,\cite{ToliasPOP2025}. This mathematical connection is valid also for interacting systems and reads as\,\cite{DornheimPRBmom,DornheimMRE2023}
\begin{equation}
{M}_{\mathrm{S}}^{(\alpha)}(\boldsymbol{q})=\frac{(-1)^{\alpha}}{\hbar^{\alpha}}\left.\frac{\partial^{\alpha}F(q,\tau)}{\partial\tau^{\alpha}}\right|_{\tau=0}\,.\label{eq:momentderivativeconnection}
\end{equation}
The most convenient expression for the non-interacting ITCF is obtained by substituting the non-interacting dynamic structure factor in the ITCF Laplace transform definition\,\cite{ToliasCTPP2025b}. We note that an equivalent expression can be obtained by substituting the analytically continued Lindhard density response to the Fourier-Matsubara series expansion of the ITCF\,\cite{ToliasJCP2024,DornheimPRB2024,DornheimEPL2024}. In normalized units where $\widetilde{M}^{(\alpha)}_{\mathrm{S}}(\boldsymbol{q})=(\hbar/E_{\mathrm{f}})^{\alpha}M^{(\alpha)}_{\mathrm{S}}(\boldsymbol{q})$, $x=q/q_{\mathrm{F}}$, $y=q'/q_{\mathrm{F}}$, the odd frequency moments of the non-interacting dynamic structure factor are given by\,\cite{ToliasPOP2025}
\begin{align}
&\widetilde{M}_{\mathrm{S}_{\mathrm{HF}}}^{(2k+1)}(x)=\frac{3\Theta}{8}x^{2k+1}\int_{0}^{+\infty}y^{2k+1}\ln{\left\{\frac{1+\exp{\left[\bar{\mu}-\frac{(x-y)^2}{4\Theta}\right]}}{1+\exp{\left[\bar{\mu}-\frac{(x+y)^2}{4\Theta}\right]}}\right\}}dy\,,\label{eq:FermioddMOM}
\end{align}
while the even frequency moments of the non-interacting dynamic structure factor are given by\,\cite{ToliasPOP2025}
\begin{align}
&\widetilde{M}_{\mathrm{S}_{\mathrm{HF}}}^{(2k)}(x)=\frac{3\Theta}{8}x^{2k}\int_{0}^{+\infty}y^{2k}\coth{\left(\frac{xy}{2\Theta}\right)}\ln{\left\{\frac{1+\exp{\left[\bar{\mu}-\frac{(x-y)^2}{4\Theta}\right]}}{1+\exp{\left[\bar{\mu}-\frac{(x+y)^2}{4\Theta}\right]}}\right\}}dy\,,\label{eq:FermievenMOM}
\end{align}
where $\bar{\mu}=\mu/T$ denotes the normalized chemical potential and where the subscript "HF" signifies the non-interacting limit (to avoid confusion with the subscript "0" that is reserved for the thermodynamic equilibrium average). It has already been pointed out that the monomial pre-factor in the integrand of Eq.(\ref{eq:FermioddMOM}) enables a very simple integration by parts regardless of the order $2k+1$, which leads to the emergence of complete Fermi integrals that correspond to $\langle\sum_i{\hat{\boldsymbol{p}}_i}^{2k}\rangle_0$ equilibrium expectation values and that the more complicated nature of the pre-factor in the integrand of Eq.(\ref{eq:FermievenMOM}) does not enable a simple integration by parts, which reflects the lack of a fundamental expression that expresses the even frequency moments of the dynamic structure factor via the equilibrium expectation value of equal time operators\,\cite{ToliasPOP2025}. The deeper mathematical investigation of the previous statement essentially led to the results reported in the present work.

\subsection{Thermodynamic averages of positive integer exponents of the momentum in the non-interacting limit}\label{sec:backgroumdMomentum}

In the paramagnetic non-interacting case of equal spin-up and spin-down electrons, the expectation value of arbitrary non-negative integer exponents $\alpha\geq0$ of the momentum magnitude is given by
\begin{equation*}
\langle{p}^{\alpha}\rangle_{\mathrm{FD}}=\frac{2}{n}\int\frac{d^3q}{(2\pi)^3}\frac{\hbar^{\alpha}q^{\alpha}}{\exp{\left(\frac{\hbar^2q^2}{2mT}-\bar{\mu}\right)}+1}\,,
\end{equation*}
where we introduced the Fermi distribution. Spherical coordinates are employed as usual with the integrations over the azimuthal angle and the inclination angle being trivial, while the normalized wavenumber $q=q_{\mathrm{F}}y$, the Fermi energy $E_{\mathrm{F}}=\hbar^2q_{\mathrm{F}}^2/2m$ and the degeneracy parameter $\Theta=T/E_{\mathrm{f}}$ are substituted for. Courtesy of $q_{\mathrm{F}}^{3}=3\pi^2n$ the density cancels out in the pre-factor leading to
\begin{equation*}
\langle{p}^{\alpha}\rangle_{\mathrm{FD}}=3\hbar^{\alpha}q_{\mathrm{F}}^{\alpha}\int_0^{\infty}dy\frac{y^{\alpha+2}}{\exp{\left(\frac{y^2}{\Theta}-\bar{\mu}\right)}+1}\,.
\end{equation*}
The standard change of variables $z=y^2/\Theta$ and the introduction of the Fermi momentum ${p}_{\mathrm{F}}=\hbar{q}_{\mathrm{F}}$ yield
\begin{equation*}
\langle{p}^{\alpha}\rangle_{\mathrm{FD}}=\frac{3}{2}{p}_{\mathrm{F}}^{\alpha}\Theta^{(\alpha+3)/2}\int_0^{\infty}dz\frac{z^{(\alpha+1)/2}}{\exp{\left(z-\bar{\mu}\right)}+1}\,.
\end{equation*}
Substituting for the tabulated complete Fermi integral, one ends up with the compact expression
\begin{equation}
\langle{p}^{\alpha}\rangle_{\mathrm{FD}}=\frac{3}{2}{p}_{\mathrm{F}}^{\alpha}\Theta^{(\alpha+3)/2}I_{(\alpha+1)/2}(\bar{\mu})\,.\label{eq:Fermimomentumavg}
\end{equation}
It is reminded that for $\alpha=0$ one has $\langle{p}^{0}\rangle_{\mathrm{FD}}=\langle1\rangle_{\mathrm{FD}}=1$, which is the normalization condition for the Fermi distribution that determines the normalized chemical potential $\bar{\mu}$.

\section{Theoretical results}\label{sec:results}

\subsection{Ansatz-based calculations for arbitrary $k$}\label{sec:resultsAnsatz}

\indent The starting point is the expression for the odd dynamic structure factor frequency moments of arbitrary order for the non-interacting gas, see Eq.(\ref{eq:FermioddMOM}). The calculations first aim to generate the Fermi distribution within the integrand.
This is achieved by decomposing into two integrals with the aid of the basic logarithmic property and by making the variable change $y\to-y$ so that the integration range becomes $(-\infty,+\infty)$
\begin{align*}
\widetilde{M}_{\mathrm{S}_{\mathrm{HF}}}^{(2k+1)}(x)=\frac{3\Theta}{8}x^{2k+1}\int_{-\infty}^{+\infty}y^{2k+1}\ln{\left\{1+\exp{\left[\bar{\mu}-\frac{(x-y)^2}{4\Theta}\right]}\right\}}dy\,;
\end{align*}
by integrating by parts courtesy of the monomial pre-factor that is rewritten as $y^{2k+1}=(\partial/\partial{y})[y^{2k+2}/(2k+2)]$
\begin{align*}
\widetilde{M}_{\mathrm{S}_{\mathrm{HF}}}^{(2k+1)}(x)=-\frac{3}{16}x^{2k+1}\int_{-\infty}^{+\infty}\frac{y^{2k+2}}{2k+2}\frac{x-y}{\exp{\left[\frac{(x-y)^2}{4\Theta}-\bar{\mu}\right]}+1}dy\,;
\end{align*}
by making the change of variables $w=(x-y)/2$ in view of the argument of the Fermi distribution kernel
\begin{align*}
\widetilde{M}_{\mathrm{S}_{\mathrm{HF}}}^{(2k+1)}(x)=-\frac{3}{4}x^{2k+1}\int_{-\infty}^{+\infty}\frac{(x-2w)^{2k+2}}{2k+2}\frac{w}{\exp{\left[\frac{w^2}{\Theta}-\bar{\mu}\right]}+1}dw\,.
\end{align*}

The calculations now aim to substitute the integral with averages of powers of the momentum over the Fermi distribution. This is achieved by applying the binomial expansion for the $(x-2w)^{2k+2}$ pre-factor
\begin{align*}
\widetilde{M}_{\mathrm{S}_{\mathrm{HF}}}^{(2k+1)}(x)&=-\frac{3}{4}\frac{x^{4k+3}}{2k+2}\int_{-\infty}^{+\infty}\sum_{j=0}^{2k+2}\binom{2k+2}{j}\left(-2\frac{w}{x}\right)^{j}\times\nonumber\\\quad&\frac{wdw}{\exp{\left[\frac{w^2}{\Theta}-\bar{\mu}\right]}+1}\,;
\end{align*}
by interchanging the integral and summation operators
\begin{align*}
\widetilde{M}_{\mathrm{S}_{\mathrm{HF}}}^{(2k+1)}(x)&=-\frac{3}{4}\frac{x^{4k+3}}{2k+2}\sum_{j=0}^{2k+2}\binom{2k+2}{j}\left(-\frac{2}{x}\right)^{j}\times\nonumber\\\quad&\int_{-\infty}^{+\infty}\frac{w^{j+1}dw}{\exp{\left[\frac{w^2}{\Theta}-\bar{\mu}\right]}+1}\,;
\end{align*}
by exploiting the parity properties of the integrand in view of the even integration boundary, which imply that the integral is non-zero only when $j$ is an odd number leading to the change of index variable $j=2i+1$ and a re-evaluation of the binomial summation limits
\begin{align*}
\widetilde{M}_{\mathrm{S}_{\mathrm{HF}}}^{(2k+1)}(x)=-\frac{3}{4}\frac{x^{4k+3}}{2k+2}\sum_{i=0}^{k}\binom{2k+2}{2i+1}\left(-\frac{2}{x}\right)^{2i+1}\int_{-\infty}^{+\infty}\frac{w^{2i+2}dw}{\exp{\left[\frac{w^2}{\Theta}-\bar{\mu}\right]}+1}\,;
\end{align*}
by restricting the integration boundary to the positive domain courtesy of the even parity of the integrand
\begin{align*}
\widetilde{M}_{\mathrm{S}_{\mathrm{HF}}}^{(2k+1)}(x)=\frac{3}{2}\frac{x^{4k+3}}{2k+2}\sum_{i=0}^{k}\binom{2k+2}{2i+1}\left(\frac{2}{x}\right)^{2i+1}\int_{0}^{+\infty}\frac{w^{2i+2}dw}{\exp{\left[\frac{w^2}{\Theta}-\bar{\mu}\right]}+1}\,;
\end{align*}
by making the change of variables $z=w^2/\Theta$ and introducing the complete Fermi integral
\begin{align*}
\widetilde{M}_{\mathrm{S}_{\mathrm{HF}}}^{(2k+1)}(x)=\frac{3}{4}\frac{x^{4k+3}}{2k+2}\sum_{i=0}^{k}\binom{2k+2}{2i+1}\Theta^{(2i+3)/2}\left(\frac{2}{x}\right)^{2i+1}I_{(2i+1)/2}(\bar{\mu})\,;
\end{align*}
by substituting for the averages of even powers of the momentum over the Fermi distribution, see Eq.(\ref{eq:Fermimomentumavg}),
\begin{equation}
\widetilde{M}_{\mathrm{S}_{\mathrm{HF}}}^{(2k+1)}(x)=\frac{1}{2}\frac{x^{4k+3}}{2k+2}\sum_{i=0}^{k}\binom{2k+2}{2i+1}\left(\frac{2}{x}\right)^{2i+1}\frac{\langle{p}^{2i}\rangle_{\mathrm{FD}}}{p_{\mathrm{F}}^{2i}}\,.\label{eq:Fermiimplicitnorm}
\end{equation}

Finally, Eq.(\ref{eq:Fermiimplicitnorm}) can be rewritten in physical units instead of normalized units using $M_{S_0}^{(2k+1)}(x)=(E_{\mathrm{f}}/\hbar)^{2k+1}\widetilde{M}_{S_0}^{(2k+1)}(x)$, $y=q/q_{\mathrm{F}}$, $p_{\mathrm{F}}=\hbar{q}_{\mathrm{F}}$, $E_{\mathrm{f}}=\hbar^2q_{\mathrm{F}}^2/(2m)$, which leads to
\begin{equation}
M_{\mathrm{S}_{\mathrm{HF}}}^{(2k+1)}(q)=\left(\frac{\hbar{q}^2}{2m}\right)^{2k+1}\frac{1}{2k+2}\sum_{i=0}^{k}\binom{2k+2}{2i+1}\left(\frac{2}{\hbar{q}}\right)^{2i}\langle{p}^{2i}\rangle_{\mathrm{FD}}\,.\label{eq:Fermiimplicitphysical}
\end{equation}
Ultimately, in Eq.(\ref{eq:Fermiimplicitphysical}), the odd dynamic structure factor frequency moments of arbitrary order for the non-interacting electron gas have been expressed via a finite summation featuring averages of even powers of the momentum over the Fermi distribution.

At this point, we remind the reader that, owing to the fact that the momentum and position operators do not commute, quantum systems have a thermodynamic equilibrium momentum distribution that depends on the strength of the interactions\,\cite{MilitzerPRL2002,HungerPRE2021,Dornheim_PRB_nk_2021,DornheimPRE2021,ToliasCTPP2025,dornheim2025applicationsphericallyaveragedpair}. We now formulate the Ansatz that the above expression for the odd dynamic structure factor frequency moments of arbitrary order for the non-interacting Fermi gas remains the same for the pure kinetic contribution to the odd dynamic structure factor frequency moments of arbitrary order for the interacting uniform electron gas. The only difference is that the averages of the even powers of the momentum are now with respect to the exact momentum distribution at thermodynamic equilibrium (which is not known analytically with the exception of certain asymptotic limits\,\cite{KimballJPA1975,YasuharaPA1976,HofmannPRB2013,RajagopalPRB1978}) and not with respect to the Fermi distribution. Therefore, our conjecture states that the following result holds for the interacting uniform electron gas
\begin{equation}
M_{\mathrm{S,KIN}}^{(2k+1)}(q)=\left(\frac{\hbar{q}^2}{2m}\right)^{2k+1}\frac{1}{2k+2}\sum_{i=0}^{k}\binom{2k+2}{2i+1}\left(\frac{2}{\hbar{q}}\right)^{2i}\langle{p}^{2i}\rangle_{0}\,,\label{eq:generalimplicitphysical}
\end{equation}
where $k$ is an arbitrary non-negative integer. In view of Eq.(\ref{eq:sumruleDSFkin}), Eq.(\ref{eq:generalimplicitphysical}) can be rewritten as
\begin{align}
&\left\langle\left[\left[\cdots\left[\left[\hat{\rho}(\boldsymbol{q}),\hat{H}\right],\hat{K}\right]\cdots\right],\left[\cdots\left[\left[\hat{\rho}^{\dagger}(\boldsymbol{q}),\hat{H}\right],\hat{K}\right]\cdots\right]\right]\right\rangle_0=\nonumber\\&\frac{(-1)^{\ell}N}{k+1}\left(\frac{\hbar^2{q}^2}{2m}\right)^{2k+1}\sum_{i=0}^{k}\binom{2k+2}{2i+1}\left(\frac{2}{\hbar{q}}\right)^{2i}\langle{p}^{2i}\rangle_{0}\,,\label{eq:generalimplicitphysicalREFORM}
\end{align}
Essentially, the $\langle{p}^{2i}\rangle_{\mathrm{FD}}\to\langle{p}^{2i}\rangle_{0}$ correspondence, which summarizes the ansatz, allowed the evaluation of an iterated commutator that features $2k+1$ kinetic operator nests, without carrying out any operator algebra.

\subsection{Explicit calculations for $k=0,1,2,3$}\label{sec:resultsexplicit}

The ansatz arises rather naturally, but it is quite bold. Thus, the validity of Eq.(\ref{eq:generalimplicitphysicalREFORM}) should still be verified with explicit calculations. Such calculations have been carried out for $k=0$ (see the f-sum rule), for $k=1$ (see the cubic sum rule), for $k=2$ and for $k=3$. Contrary to popular belief, in our experience, nested commutator evaluations based on the second quantization tend to be more complicated than nested commutator evaluations based on the canonical commutation relation. Hence, the explicit calculations are based on substituting for the microscopic density operator $\hat{\rho}(\boldsymbol{q})=\sum_ie^{-\imath\boldsymbol{q}\cdot\hat{\boldsymbol{r}}_i}$ and its Hermitian conjugate $\hat{\rho}^{\dagger}(\boldsymbol{q})=\sum_ie^{+\imath\boldsymbol{q}\cdot\hat{\boldsymbol{r}}_i}$, substituting for the kinetic energy operator $\hat{K}=\sum_i\hat{\boldsymbol{p}}_i^2/(2m)$ and employing the canonical commutator $[\hat{r}_{i\mu},\hat{p}_{j\lambda}]=\imath\hbar\delta_{ij}\delta_{\lambda\mu}$, where $i,j$ are particle labels and $\lambda,\mu$ are vector component labels. In particular, the following corollaries have been employed that can be derived with standard commutator identities,
\begin{align*}
&\left[e^{-\imath\boldsymbol{q}\cdot\hat{\boldsymbol{r}}_i},\left(\boldsymbol{q}\cdot\hat{\boldsymbol{p}}_j\right)\right]=\hbar{q}^2e^{-\imath\boldsymbol{q}\cdot\hat{\boldsymbol{r}}_i}\delta_{ij}\,,\\
&\left[e^{-\imath\boldsymbol{q}\cdot\hat{\boldsymbol{r}}_i},\left(\boldsymbol{q}\cdot\hat{\boldsymbol{p}}_j\right)^2\right]=2\hbar{q^2}\left\{\left(\boldsymbol{q}\cdot\hat{\boldsymbol{p}}_i\right)+\frac{1}{2}\hbar{q}^2\right\}e^{-\imath\boldsymbol{q}\cdot\hat{\boldsymbol{r}}_i}\delta_{ij}\,,\nonumber\\
&\left[e^{-\imath\boldsymbol{q}\cdot\hat{\boldsymbol{r}}_i},\left(\boldsymbol{q}\cdot\hat{\boldsymbol{p}}_j\right)^3\right]=3\hbar{q}^2\left\{\left(\boldsymbol{q}\cdot\hat{\boldsymbol{p}}_i\right)^2+\hbar{q}^2\left(\boldsymbol{q}\cdot\hat{\boldsymbol{p}}_i\right)\right.\nonumber\\&\qquad\qquad\qquad\qquad\qquad\qquad\qquad\left.+\frac{1}{3}\hbar^2{q}^4\right\}e^{-\imath\boldsymbol{q}\cdot\hat{\boldsymbol{r}}_i}\delta_{ij}\,,\nonumber\\
&\left[e^{-\imath\boldsymbol{q}\cdot\hat{\boldsymbol{r}}_i},\left(\boldsymbol{q}\cdot\hat{\boldsymbol{p}}_j\right)^4\right]=4\hbar{q}^2\left\{\left(\boldsymbol{q}\cdot\hat{\boldsymbol{p}}_i\right)^3+\frac{3}{2}\hbar{q}^2\left(\boldsymbol{q}\cdot\hat{\boldsymbol{p}}_i\right)^2\right.\nonumber\\&\qquad\qquad\qquad\qquad\qquad\left.+\hbar^2{q}^4\left(\boldsymbol{q}\cdot\hat{\boldsymbol{p}}_i\right)+\frac{1}{4}\hbar^3{q}^6\right\}e^{-\imath\boldsymbol{q}\cdot\hat{\boldsymbol{r}}_i}\delta_{ij}\,.\nonumber
\end{align*}
In addition, the following corollary has been employed that can be elegantly derived with mathematical induction,
\begin{align*}
\left[\left[\cdots\left[\left[\hat{\rho}(\boldsymbol{q}),\hat{H}\right],\hat{K}\right]\cdots,\hat{K},\hat{K}\right]\right]&=\frac{\hbar^{\ell}}{m^{\ell}}\sum_{i}\left(\boldsymbol{q}\cdot\hat{\boldsymbol{p}}_i+\frac{1}{2}\hbar{q}^2\right)^{\ell}e^{-\imath\boldsymbol{q}\cdot\hat{\boldsymbol{r}}_i}\,,\nonumber
\end{align*}
where $\ell$ is the number of $\hat{\rho}(\boldsymbol{q})-$bound operators.

The kinetic contribution to the first frequency moment sum rule ($k=0$) is evaluated with the choice $\ell=0$ for the distribution of the nested commutators, the application of the fifth corollary for $\ell=1$ and the application of the first corollary. The above lead to the $k=0$ kinetic contribution of
\begin{equation}
\left\langle\left[\left[\hat{\rho}(\boldsymbol{q}),\hat{H}\right],\hat{\rho}^{\dagger}(\boldsymbol{q})\right]\right\rangle_0=N\frac{\hbar^2{q}^2}{m}\,.\label{eq:explicit1}
\end{equation}
The result of the explicit calculation coincides with the entire f-sum rule (since the kinetic contribution is the only contribution for this frequency moment order), see Eq.(\ref{eq:DSFMOM1}), and the result of the ansatz-based calculation, see Eq.(\ref{eq:generalimplicitphysicalREFORM}), for $k=0$.

The kinetic contribution to the third frequency moment sum rule ($k=1$) is evaluated with the choice $\ell=1$ for the distribution of the nested commutators, the application of the fifth corollary for $\ell=2$, the application of the fifth corollary for $\ell=1$ and $\boldsymbol{q}\to-\boldsymbol{q}$, the application of the first two corollaries and the aid of the equilibrium expression $\langle\sum_{i}(\boldsymbol{q}\cdot\hat{\boldsymbol{p}}_i)^2/2m\rangle_0=(1/3)Nq^2\langle{K}\rangle_0$. The above lead to the $k=1$ kinetic contribution of
\begin{align}
&\left\langle\left[\left[\left[\hat{\rho}(\boldsymbol{q}),\hat{H}\right],\hat{K}\right],\left[\hat{\rho}^{\dagger}(\boldsymbol{q}),\hat{H}\right]\right]\right\rangle_0=-N\frac{\hbar^6{q}^6}{4m^3}\nonumber\\&\qquad-2N\frac{\hbar^4{q}^4}{m^2}\langle{K}\rangle_0\,.\label{eq:explicit3}
\end{align}
The result of the explicit calculation coincides with the kinetic contribution to the cubic sum rule, that is represented by the first two adders of Eq.(\ref{eq:DSFMOM3}), and the result of the ansatz-based calculation, see Eq.(\ref{eq:generalimplicitphysicalREFORM}), for $k=1$.

The kinetic contribution to the fifth frequency moment sum rule ($k=2$) is evaluated with the choice $\ell=2$ for the distribution of the nested commutators, the application of the fifth corollary for $\ell=3$, the application of the fifth corollary for $\ell=2$ and $\boldsymbol{q}\to-\boldsymbol{q}$, the application of the first three corollaries and the aid of the equilibrium expressions $\langle\sum_{i}(\boldsymbol{q}\cdot\hat{\boldsymbol{p}}_i)^2/2m\rangle_0=(1/3)Nq^2\langle{K}\rangle_0$, $\langle\sum_{i}(\boldsymbol{q}\cdot\hat{\boldsymbol{p}}_i)^4/4m^2\rangle_0=(1/5)Nq^4\langle{K^2}\rangle_0$. The above lead to the $k=2$ kinetic contribution of
\begin{align}
&\left\langle\left[\left[\left[\left[\hat{\rho}(\boldsymbol{q}),\hat{H}\right],\hat{K}\right],\hat{K}\right],\left[\left[\hat{\rho}^{\dagger}(\boldsymbol{q}),\hat{H}\right],\hat{K}\right]\right]\right\rangle_0={N}\frac{\hbar^{10}{q}^{10}}{16m^5}\nonumber\\&\qquad+\frac{5}{3}N\frac{\hbar^8q^8}{m^4}\langle{K}\rangle_0+4N\frac{\hbar^6{q}^6}{m^3}\langle{K^2}\rangle_0\,.\label{eq:explicit5}
\end{align}
The result of the explicit calculation coincides with the result of the ansatz-based calculation, see Eq.(\ref{eq:generalimplicitphysicalREFORM}), for $k=2$.

The kinetic contribution to the seventh frequency moment sum rule ($k=3$) is evaluated with the choice $\ell=3$ for the distribution of the nested commutators, the application of the fifth corollary for $\ell=4$, the application of the fifth corollary for $\ell=3$ and $\boldsymbol{q}\to-\boldsymbol{q}$, the application of the first four corollaries and the aid of the equilibrium expressions $\langle\sum_{i}(\boldsymbol{q}\cdot\hat{\boldsymbol{p}}_i)^2/2m\rangle_0=(1/3)Nq^2\langle{K}\rangle_0$, $\langle\sum_{i}(\boldsymbol{q}\cdot\hat{\boldsymbol{p}}_i)^4/4m^2\rangle_0=(1/5)Nq^4\langle{K^2}\rangle_0$ and $\langle\sum_{i}(\boldsymbol{q}\cdot\hat{\boldsymbol{p}}_i)^6/8m^3\rangle_0=(1/7)Nq^6\langle{K^3}\rangle_0$. The above lead to the $k=3$ kinetic contribution of
\begin{align}
&\left\langle\left[\left[\left[\left[\left[\hat{\rho}(\boldsymbol{q}),\hat{H}\right],\hat{K}\right],\hat{K}\right],\hat{K}\right],\left[\left[\left[\hat{\rho}^{\dagger}(\boldsymbol{q}),\hat{H}\right],\hat{K}\right],\hat{K}\right]\right]\right\rangle_0=\nonumber\\&\qquad-N\frac{\hbar^{14}q^{14}}{64m^7}-7N\frac{\hbar^{12}q^{12}}{8m^6}\langle{K}\rangle_0-7N\frac{\hbar^{10}q^{10}}{m^5}\langle{K}^2\rangle_0\nonumber\\&\qquad-8N\frac{\hbar^8q^8}{m^4}\langle{K}^3\rangle_0\,.\label{eq:explicit7}
\end{align}
The result of the explicit calculation coincides with the result of the ansatz-based calculation, see Eq.(\ref{eq:generalimplicitphysicalREFORM}), for $k=3$.

\section{Summary and discussion}\label{outro}

An exact expression has been derived for the kinetic contribution to the odd (arbitrary order) frequency moments of the dynamic structure factor. The expression comprises a finite summation that features the averages of even (all lower orders) powers of the momentum over the exact momentum distribution. The derivation is based on analytical expressions that are only available for the non-interacting Fermi gas with the transition to the general interacting case based on the ansatz that thermodynamic averages over the Fermi distribution can be substituted with thermodynamic averages over the exact momentum distribution. The ansatz has been successfully validated against known results (for the first and third frequency moments) and against new explicit calculations based on the canonical commutator relation and the fundamental nested commutator formula of linear response theory (for the fifth and seventh frequency moments).

The pure kinetic contribution to the arbitrary odd order frequency moments of the dynamic structure factor constitutes one of many contributions to the overall frequency moment. It is possible to check whether it is the dominant contribution utilizing the recent framework of Dornheim \emph{et al.} that allows the direct extraction of arbitrary order frequency moments of the dynamic structure factor from QMC simulations\,\cite{DornheimPRBmom}. In case this holds in some phase diagram regions, then the present exact expression can be exploited to formulate advanced variants of the non-perturbative sum rule approach to dynamic properties\,\cite{FilinovPRB2023}, construct accurate approximations of the even order frequency moments of the dynamic structure factor via a recently derived infinite series representation\,\cite{ToliasPOP2025}, obtain novel convenient QMC estimators of the averages of even exponents of the momentum over the exact momentum distribution and formulate additional constraints that will enable reliable extraction of dynamic structure factors from QMC simulations (see the analytic continuation\,\cite{epstein2008badtruth,JARRELL1996133,chuna2025dual} problem).

Even though this theoretical investigation was motivated by the finite temperature uniform electron gas, the general $2k+1$ order frequency moment expression of Eq.(\ref{eq:generalimplicitphysical}) is valid for any non-relativistic three-dimensional paramagnetic one-component fermionic system.
The expression exclusively concerns the kinetic contribution, thus the type of the interaction and the strength of the correlations do not appear explicitly. However, they implicitly influence the results through their effect on the exact momentum distribution. It is worth noting that our expression can be readily extended for arbitrary system dimensionality, arbitrary spin polarization and fermionic mixtures. Regardless of the dimensionality, spin polarization and number of species, it is expected that the mathematical structure of the finite summation, which features the averages of even ($0$ to $2k$) exponents of the momentum over the exact momentum distribution, will persist but that the summation coefficients and pre-factors will differ. In the same spirit, similar expressions are expected to hold for any spin statistics (bosons, boltzmannons, $q-$ons), but the veracity of this statement requires a dedicated investigation.

\section*{Acknowledgements}

This work was partially supported by the Center for Advanced Systems Understanding (CASUS), financed by Germany's Federal Ministry of Education and Research (BMBF) and the Saxon state government out of the State budget approved by the Saxon State Parliament. This work has received funding from the European Research Council (ERC) under the European Union's Horizon 2022 research and innovation programme (Grant agreement No. 101076233, "PREXTREME"). Views and opinions expressed are however those of the authors only and do not necessarily reflect those of the European Union or the European Research Council Executive Agency. Neither the European Union nor the granting authority can be held responsible for them.

\bibliography{main_bib}

\end{document}